# A General-Purpose Data Harmonization Framework: Supporting Reproducible and Scalable Data Integration in the RADx Data Hub

Jimmy K. Yu, PhD, Marcos Martínez-Romero, PhD, Matthew Horridge, PhD,
Mete U. Akdogan, PhD, and Mark A. Musen, MD, PhD
Center for Biomedical Informatics Research, Stanford University, Stanford, CA, USA

**Abstract**

*In the age of big data, it is important for primary research data to follow the FAIR principles of findability, accessibility, interoperability, and reusability. Data harmonization enhances interoperability and reusability by aligning heterogeneous data under standardized representations, benefiting both repository curators responsible for upholding data quality standards and consumers who require unified datasets. However, data harmonization is difficult in practice, requiring significant domain and technical expertise. We present a software framework to facilitate principled and reproducible harmonization protocols. Our framework implements a novel strategy of building harmonization transformations from parameterizable primitive operations, such as the assignment of numerical values to user-specified categories, with automated bookkeeping for executed transformations. We establish our data representation model and harmonization strategy and then report a proof-of-concept application in the context of the RADx Data Hub. Our framework enables data practitioners to execute transparent and reproducible harmonization protocols that align closely with their research goals.*

**Introduction**

As a response to the COVID-19 pandemic, the National Institutes of Health (NIH) established the Rapid Acceleration of Diagnostics (RADx) program both to fund primary research on novel diagnostic technologies and to better understand social determinants of health outcomes.[1] A key component of the RADx program is its centralized data platform, the RADx Data Hub, that serves as a much-needed platform for sharing primary COVID-19 research data, especially to facilitate retrospective studies on the COVID-19 response.[2] However, the use of data for secondary research poses a set of unique challenges that are well summarized by the FAIR (Findable, Accessible, Interoperable, and Reusable) principles of data stewardship:[3] 1) datasets relevant to a research goal should be searchable and easily discovered, 2) once found, the datasets should be easily obtained, 3) datasets should be easy to integrate into a unified dataset for downstream applications or storage, and 4) it should be practical to use the datasets. We recently presented work on software development, data curation, and metadata management to facilitate FAIRness in establishing the RADx Data Hub.[2] Here, we focus on the interoperability problem using a novel approach to data harmonization.

Data science applications often require a unified dataset to achieve the requisite statistical power to investigate a hypothesis or to provide sufficient training data for a computational model. Unfortunately, it can be challenging to integrate data from heterogeneous sources into a combined dataset. Datasets from different sources, even when covering the same domain and concepts, may define or represent data differently. Dataset heterogeneity is addressed by implementing data harmonization strategies. Harmonization is the practice of aligning heterogeneous datasets such that they comply with a chosen common standard. Prospective harmonization can be viewed as an ideal solution: establish data representations upfront so that all collected data share common semantics and syntax. While this is a viable strategy for establishing a research program from scratch, it can be implemented only by the original researchers or enforced by a central governing body.[4–6] Moreover, predefined data representations may be too rigid to support downstream applications or to adapt to unforeseen circumstances during data collection. Prospective harmonization is most powerful when a governing body, such as a funding agency or leadership of a research consortium, can exert control over data collection.[6] However, prospective harmonization is not an option for secondary researchers and cannot be implemented by data stewards and repository managers who are responsible for harmonizing a dataset prior to its ingestion into a data repository. Instead, retrospective (post hoc) harmonization must be performed to align heterogeneous datasets under new or existing data representations. The urgency of a rapid COVID-19 response and the diversity in research domains within the RADx programs limited the extent and effectiveness of its original prospective harmonization strategy,[2] thereby requiring additional post hoc harmonization to improve the alignment of ingested data to standards expanded and refined after the fact.

The implementation of a harmonization strategy requires domain expertise and is inherently subjective. A primary researcher may prioritize a representation for initial data collection that is highly specific to their research goals, the steward of a data repository may prioritize a representation that maximizes homogeneity across many related research studies, and a secondary researcher may be interested in a data representation that allows integration with data obtained from another source. For example, the managers of a data repository may require submitted data to strictly conform to a taxonomy or ontology, whereas a data scientist may have a higher tolerance for grouping concepts that are approximately equivalent. We recognize that there is no universal solution to data harmonization. Accordingly, we present a new framework for data harmonization that supports ad hoc harmonization protocols by placing emphasis on rigorously defined, reproducible, and transparent data transformations.

While there is an extensive literature on data integration and harmonization techniques in biomedical informatics that focus on best practices, guidelines, and methodology,[6–8] in practice, retrospective harmonization has favored the use of task-specific scripts to massage data into their desired representations, limiting their flexibility, adaptability, and transparency.[9–13] Similarly, other software tools exist to facilitate harmonization,[13–16] but their specialization to specific domains, focus on repository management, or complex infrastructure limits their general applicability. Our framework allows data practitioners across the spectrum to implement potentially complex data manipulations that draw from their expertise and conform to their research goals. We introduce a reproducible strategy that builds data transformations from primitive operations and executes them with automated record keeping for validation, reproduction, and later use. This work was done in support of and inspired by the needs of the RADx Data Hub and, as a result, focuses on applications to tabular biomedical data. However, we have designed our framework to be extensible to general purposes in data management.

**Methods**

First, we define our data representation model with a focus on the involved entities and their relationships. This model generalizes the one used by the RADx Data Hub.[2] We then define the idea of a *harmonization rule* in our framework and detail our strategy for implementing principled and reproducible data transformations.

*Data Representation Model*

Our data representation model has four key entities: the *variable*, the *data element*, the *data dictionary*, and the *data file*. Together, they define how data are conceptualized, structured, and represented. Their relationships are introduced in Figure 1 and will be described in depth in the remainder of the section.

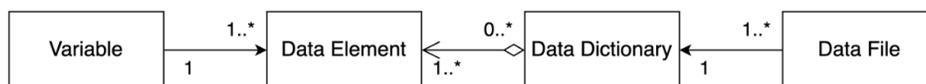

**Figure 1.** Unified Modeling Language (UML) diagram of our data representation model. A variable can be implemented by one or more data elements, data dictionaries aggregate multiple data elements, and a data file corresponds to exactly one data dictionary, whereas the same data dictionary can be used for multiple data files.

The *variable* in our model represents an abstract concept for which values are recorded during data collection. The *data element* entity represents its practical implementation and encodes the computational aspects needed to represent, record, and store collected data. The *data element* in our model generalizes the more familiar notion of the common data element (CDE).[17,18] A CDE is a precisely defined specification of a question and its allowed answers, for example, by its computational data type or an enumerated set of choices. CDEs are used to standardize the representation of data, such as in a curated repository. The *data element* in our model includes any representation without the expectation that it adheres to a common or predefined standard. To make this concrete, we use an example of a population health survey that asks for the age of a survey participant as one of its questions. The survey participant sees the prompt "What is your age?" with a free response text box to enter a value. For this question, the "age" concept is the *variable*, whereas the prompt and the data type of a text response comprises its implementation as a *data element*.

The specification of a *data element* is subjective; a different researcher could choose an alternative implementation for "age" to better suit their needs. For example, if an age range is preferred as a means of anonymizing survey responses, the survey could ask the same prompt but instead enumerate multiple-choice options with year ranges for the age of the participant, as demonstrated in Figure 2. The same *variable* for "age" can be implemented in different ways by different data elements to achieve varying degrees of precision or specificity, as indicated in Figure 1. Even a small difference in *data elements* for the same *variable* results in heterogeneity that must be resolved as part of harmonization.

| Concept | Data Element | Prompt | Response Type |
|---|---|---|---|
| age | age_text | What is your age? | String |
| age | age_range | What is your age? | Categorical:<br>30 or Under<br>31-40<br>41-50<br>51-60<br>61-70<br>Over 70 |

**Figure 2.** Two *data element* implementations of the "age" *variable*. Each *data element* is defined in its own row. The top *data element* implements the "age" *variable* by asking for a text response. The bottom *data element* implements the "age" *variable* by requesting a multiple choice response for categorical age ranges.

The *data file* is the entity in our model that stores instances of collected data. A *data file* obeys a schema defined by a *data dictionary*. We have adopted the standard notion of the *data dictionary* in data management as a metadata entity that specifies the semantics for recorded data. A *data dictionary* in our model aggregates one or more *data elements* to define the data collection schema (Figure 1). A *data file* must have exactly one *data dictionary*, whereas the same *data dictionary* may be used as the schema for multiple *data files* (Figure 1), such as when a research group conducts the same experiment as part of different campaigns or when multiple research groups use a shared data collection template. Using the example of the population health survey, each question would be represented by a data element. As shown in Figure 3, a *data dictionary* contains the set of all *data elements* implemented as part of the survey and the *data file* records the responses provided by the study participants. Although the name *data file* implies document-based storage, the concept can be abstracted to cover any data storage, such as a database table that adheres to the schema defined by the *data dictionary*.

**Data Dictionary**

| Concept | Data Element | Prompt | Response Type |
|---|---|---|---|
| age | age_text | What is your age? | String |
| sex assigned at birth | sex | What is your biological sex assigned at birth? | Coded Values:<br>0, Female<br>1, Male<br>2, Intersex<br>3, Prefer not to answer |
| COVID-19 vaccination status | cov19_vaccination_status | Are you vaccinated against COVID-19? | Coded Values:<br>0, Yes<br>1, No<br>2, Do not know<br>3, Prefer not to answer |

**Data File**

| record_id | age_text | sex | cov19_vaccination_status |
|---|---|---|---|
| 1 | 23 | 1 | 0 |
| 2 | 47 | 1 | 0 |
| 3 | 31 | 0 | 0 |
| 4 | 56 | 1 | 1 |
| 5 | 23 | 0 | 0 |
| 6 | 45 | 0 | 3 |
| 7 | 68 | 3 | 0 |
| 8 | 25 | 1 | 0 |
| 9 | 34 | 1 | 1 |
| 10 | 93 | 0 | 0 |

**Figure 3.** The left side shows an example *data dictionary* that includes three *data elements*: "age_text," "sex," and "cov19_vaccination_status." The right side shows an example *data file* corresponding to this *data dictionary*, with instances for each *data element* stored in a column and one record per row. Values for the "age_text" *data element* are recorded in text, while "sex" and "cov19_vaccination_status" store integer values that correspond to the coded values specified by the data dictionary.

*Harmonization Rules*

Suppose we obtain a data file that must undergo harmonization to a common data representation. This common representation may be defined independently by a governing body (of a data repository, for example) or ad hoc by a data consumer as part of an analysis or mining task. For the former case, it may be desirable to enforce a standardized representation by requiring the use of a set of vetted CDEs.[17,18] The goal then becomes to align the data file to these CDEs. For harmonization applications that do not seek to standardize data for long term storage, it may be sufficient to pick a set of data elements that are easy to work with as the common representation.

In order to systematically resolve heterogeneities in data sets, we introduce the *harmonization rule* as a structure that defines the participants and operations that are required to transform data during harmonization. A *harmonization rule* is characterized by three components: 1) a source data element, 2) a target data element, and 3) a *mapping function* that transforms an instance of the source to an instance of the target, as depicted in Figure 4. Source data elements are drawn directly from the original data file and its data dictionary, as they define the representation of data as intended by the primary researchers who produced the data. When a standard data representation already exists, as in the case of a data repository, the target data elements are the pre-established CDEs. For ad hoc data analysis, a data practitioner may wish to define new target data elements out of convenience or simply borrow a data element from one of the ingested datasets to use as a common standard. Source and target data elements that represent equivalent or similar concepts are good candidates for harmonization.[19,20] A data practitioner may also demand higher rigor and require precise entity alignment between source and target data elements.[21–23]

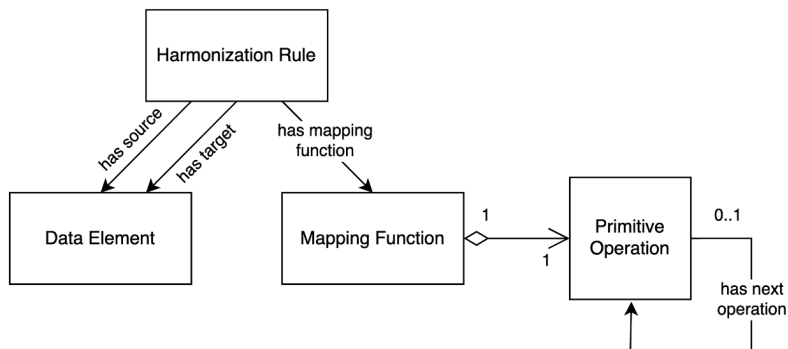

**Figure 4.** Anatomy of a harmonization rule. A harmonization rule is composed of a source data element, a target data element, and a mapping function that transforms values of the source into values of the target. The mapping function is constructed from one or more primitive operations to be applied serially.

After source and target data elements are identified, data under the source data element representation can be transformed into instances that obey the target representation. Let $V_1$ be the set of values belonging to the source data element and $V_2$ be the set of values belonging to the target data element. A mapping function that transforms $V_1$ into $V_2$ is defined as $h: V_1 \rightarrow V_2$ such that for a value $x \in V_1$, $h(x) = y$ where $y \in V_2$. For example, to convert a distance value from units of kilometers to units of miles, a value $x$ measured in kilometers would be passed as an input into the unit conversion function $h(x)$ to output the equivalent distance $y$ in miles. In practice, mapping functions are often bespoke, hand-crafted rules that can be applied manually or with the support of automation by simple scripts.[9–13]

Reporting such data transformations requires meticulous record keeping, precise language, and published code. Even with all three, reproducing the harmonization protocol may be challenging if the written protocols are lacking in sufficient detail or the published code is missing manual or implicit steps. This situation motivates our approach to structure data transformations in harmonization rigorously by leveraging atomic *primitive operations* as building blocks for *mapping functions*. This bottom-up approach allows for highly transparent, traceable, and reproducible harmonization transformations, as will be demonstrated by example in the **Results** section later.

*Primitive Operations*
We propose that a harmonization mapping function can be expressed as a series of elementary operations (referenced herein as "primitives") applied serially. We will guide the discussion of primitives and their role in harmonization by example in Figure 5, which extends the previous health survey example by walking through the harmonization process that transforms data collected using the "age_text" data element into instances of the "age_range" data element.

Our initial design features nine primitives, each of which can be parameterized to refine its behavior, as summarized in Table 1. The framework can easily be extended to accommodate additional primitives as needed. Primitives serve as elementary building blocks from which more complex harmonization mapping functions can be composed. Given $G$ as the set of primitive operations, a primitive operation $g \in G$ accepts a single value, $x$, as input and has a parameterization, $R$, that tunes its behavior. A mapping function $h(x)$ can then be expressed as a sequence of $n$ parameterized primitive operations with each successive primitive operating on the output of the previous:
$h(x) = g_n (g_{n-1} (...g_2(g_1(x \mid R_1) \mid R_2)...) \mid R_n)$.

Figure 5a shows the composition of a mapping function using two primitive operations. To transform a value of "age_text" into an "age_range" category, the age text must first be converted from a string data type into a numerical data type. This operation uses the "Cast" primitive, which is parameterized by the original data type, string, and the desired data type, integer. The integer age value produced by the "Cast" primitive is then sent through a second primitive, "Bin," that assigns the age value to the appropriate age range. The "Bin" primitive is parameterized by the age range categories defined by the "age_range" data element, as described earlier in Figure 2.

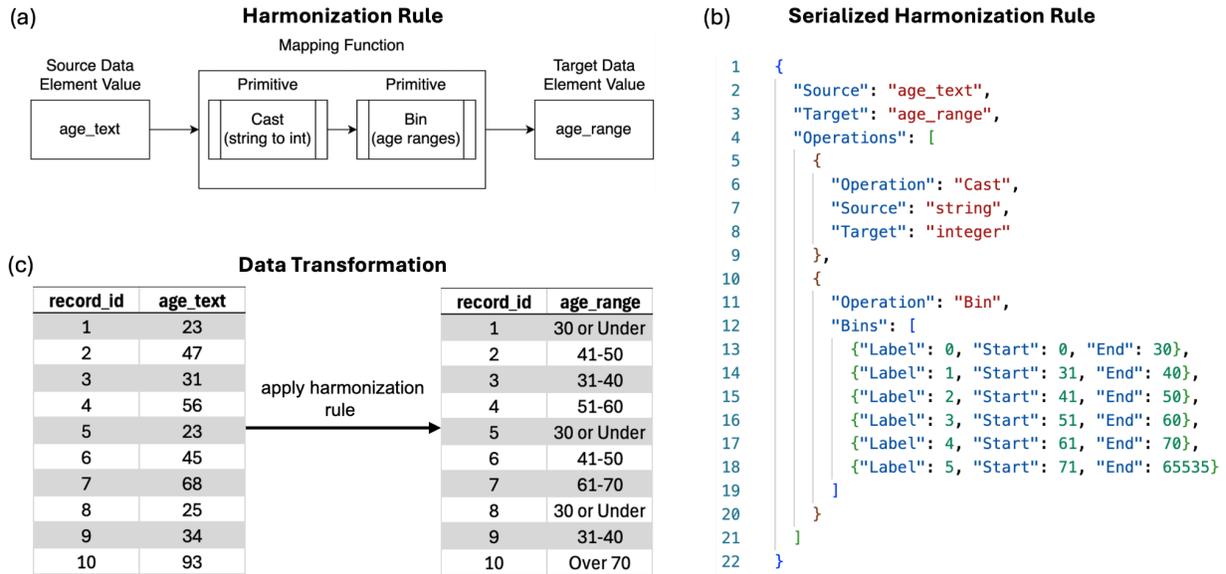

**Figure 5.** Example of a harmonization rule that transforms instances of the text-based age data element to the age range data element from Figure 2. (a) The mapping function is composed of two primitive operations, first converting string to integer and then mapping the integer age value into its appropriate category. (b) JSON serialization of the harmonization rule shown in (a). (c) Data file before and after application of the harmonization rule to transform "age_text" to "age_range."

| Primitive | Parameterization | Description | Example |
|---|---|---|---|
| ConvertUnits | source unit; target unit | convert a numerical source value from one unit of measurement to another | *ConvertUnits*(1000) → 1<br>params: {source: meters, target: kilometers} |
| Truncate | maximum allowed length | truncate a text source value using a parameterized character limit | *Truncate*("94305-2005") → "943"<br>params: {length: 3} |
| Cast | source data type; target data type | convert value of source to a target data type, e.g., string to integer | *Cast*("1") → 1<br>params: {source: string, target: integer} |
| EnumToEnum | re-coding scheme as a key–value map | map a coded source value to a coded value under another coding scheme | *EnumToEnum*(2) → 1<br>params: {mapping: 1 → 0, 2 → 1, 3 → 2} |
| Bin | labeled intervals or histogram bins | assign a numerical source value to a categorical variable based on histogram bins | *Bin*(137) → "diabetes"<br>params: {bins: [MIN, 99] → "normal",<br>[100, 125] → "prediabetes",<br>[126, MAX] → "diabetes"} |
| Reduce | operation name (sum, any, none, all, one-hot) | convert a vector source value to a single summarizing value or boolean | *Reduce*([0, 0, 1, 0]) → True<br>params: {operation: any} |
| ConvertDate | source date format; target date format | convert a date in a source format to a target format | *ConvertDate*("03/14/2025") → "2025-03-14"<br>params: {source: %m/%d/%Y,<br>target: %Y-%m-%d} |
| Round | precision (most significant decimal place) | round a decimal source value to the parameterized precision | *Round*(13.226) → 13.23<br>params: {precision: 2} |
| Threshold | lower bound; upper bound | apply a floor and ceiling function to a numerical source value | *Threshold*(16) → 10<br>params: {lower: 0, upper: 10} |

**Table 1.** Primitive functions currently implemented in our harmonization framework library. Each function takes a single input, the value of the source data instance, as an argument, but the function may be parameterized by options specific to each primitive to define its behavior. The last column provides an example application of each primitive with a hypothetical parameterization.

The purpose of constructing a predefined parameterizable set of primitive functions is to allow for the reproducible and principled construction of tailored harmonization mappings. Harmonization rules constructed using this framework avoid ambiguous text-based protocols that may assume or omit key data processing steps. The primitive functions lend themselves to serialization, which is the encoding of data in a structured and machine-readable form. Serialization enables a complete specification of a data transformation without ambiguity. Each parameterized primitive function and, by extension, a complete harmonization rule can be serialized to a set of instructions for performing the transformation, and this serialization can be stored to persistent storage for record keeping, sharing, or publication. Our framework currently supports the serialization of harmonization rules in JSON format.

The serialization of the harmonization rule that transforms "age_text" into "age_range" is shown in Figure 5b. The original data element is labeled as the "Source," the desired data element is labeled as the "Target," and the "Operations" record the serialized mapping function depicted in Figure 5a, which contains the "Cast" and "Bin" primitives in order with their parameters. Finally, Figure 5c illustrates the application of the harmonization rule described by Figure 5a,b. The input data file contains ten records of participant ages in text format using the "age_text" data element, and applying the harmonization rule produces a new, transformed data file that contains the same ten records with the text ages converted into categorical age ranges that comply with the specification of the "age_range" data element. The benefits of serializing harmonization rules are especially important in the context of destructive harmonization procedures, such as the "age_text" to "age_range" transformation, for which the original data cannot be recovered from their harmonized form. A harmonization rule built using this framework enables tailored data transformations by leveraging the flexibility to arbitrarily compose and parameterize primitive operations with well-defined and reproducible behavior.

**Software Interface and Features**

We have implemented our harmonization framework as a Python library, and its source code is openly available on GitHub (see **Code Availability**). The library provides an interface to the core entities of the data model, the primitive functions for constructing complex mapping functions, and a data structure for defining harmonization rules. In addition, the library includes a suite of utilities for executing batched data transformations with automated logging. A user can leverage the application programming interface (API) to define tailored harmonization rules and apply them to their data files programmatically. The framework currently supports this process through traditional scripting or through an interactive programming environment such as a Jupyter notebook. While the framework supports production-quality harmonization efforts, it currently requires proficiency in programming to use in its proof-of-concept state. The library interface requires a user: 1) to select a set of data element specifications as harmonization targets, 2) to manually pair source and target data elements, and 3) to define a mapping function for each source and target data element pair in terms of the primitive operations listed in Table 1. Utilities provided by the framework can then be used to automate the harmonization transformation.

Our framework places a heavy emphasis on the preservation of harmonization rules and the reproducibility of harmonization transformations. We have two primary objectives in preserving harmonization rules. First, we desire the ability to communicate and reproduce a harmonization protocol exactly, avoiding the ambiguity that a written procedure may carry. To support this functionality, all harmonization rules developed with our framework can be serialized to JSON format for storage or transmission. The harmonization rule can be deserialized at a later time or by a different party using the framework to reconstruct the harmonization rule exactly. In support of this feature, the harmonization framework provides utilities for interfacing with file-based storage or key–value data stores that support queries for harmonization rules based on the source and target data elements.

Second, all data transformations should be transparent. Harmonization is an inherently destructive process, and mappings may not be reversible. For example, the harmonization process from "age_text" to "age_range" in Figure 5 is not reversible; a numerical age value cannot be exactly recovered from an age range alone. It is therefore important to maintain a record of the harmonization rules that have been applied to a data file in the order they are performed while preserving the original data file. The framework achieves this capability by implementing detailed logging to accompany all applications of a harmonization rule. A log entry records two fields in JSON format: 1) the full serialization of the harmonization rule as an "action" and 2) the name of the dataset to which it was applied in the "dataset" field. The harmonization log enables a complete reproduction of any harmonization procedures applied to a data file as long as the data practitioner is in possession of the original data file. Without verbose and meticulous documentation, details of how the data have been manipulated may be lost, leaving data consumers without the means to recreate the aggregate dataset or verify its correctness. This outcome may ultimately prevent the verification or reproduction of analysis results and is especially significant when harmonized, cleaned, and aggregated datasets are used for analysis and then revisited at a later time.

## Results

As a proof of concept, we show a representative application of our harmonization framework to two synthetic data files generated from real data elements used by the RADx Underserved Populations (RADx-UP) and RADx Radical (RADx-rad) projects.[1] RADx-UP and RADx-rad both record data for a variable concerning the "employment status" of survey respondents. RADx-UP implements this variable as the "current_employment_status" data element, whereas RADx-rad implements it differently, as the "employment" data element. Data dictionaries that include these data elements are shown in Figure 6a for RADx-UP and in Figure 6b for RADx-rad. To also demonstrate the framework's ability to accommodate a harmonization protocol involving multiple data elements, we also include synthetic data elements for the "commute distance" concept, with the RADx-UP dataset using "commute_distance_miles" and the RADx-rad dataset using "commute_distance_km." We have generated synthetic data files from these data dictionaries to show their contents, as data in the RADx Data Hub have controlled access.

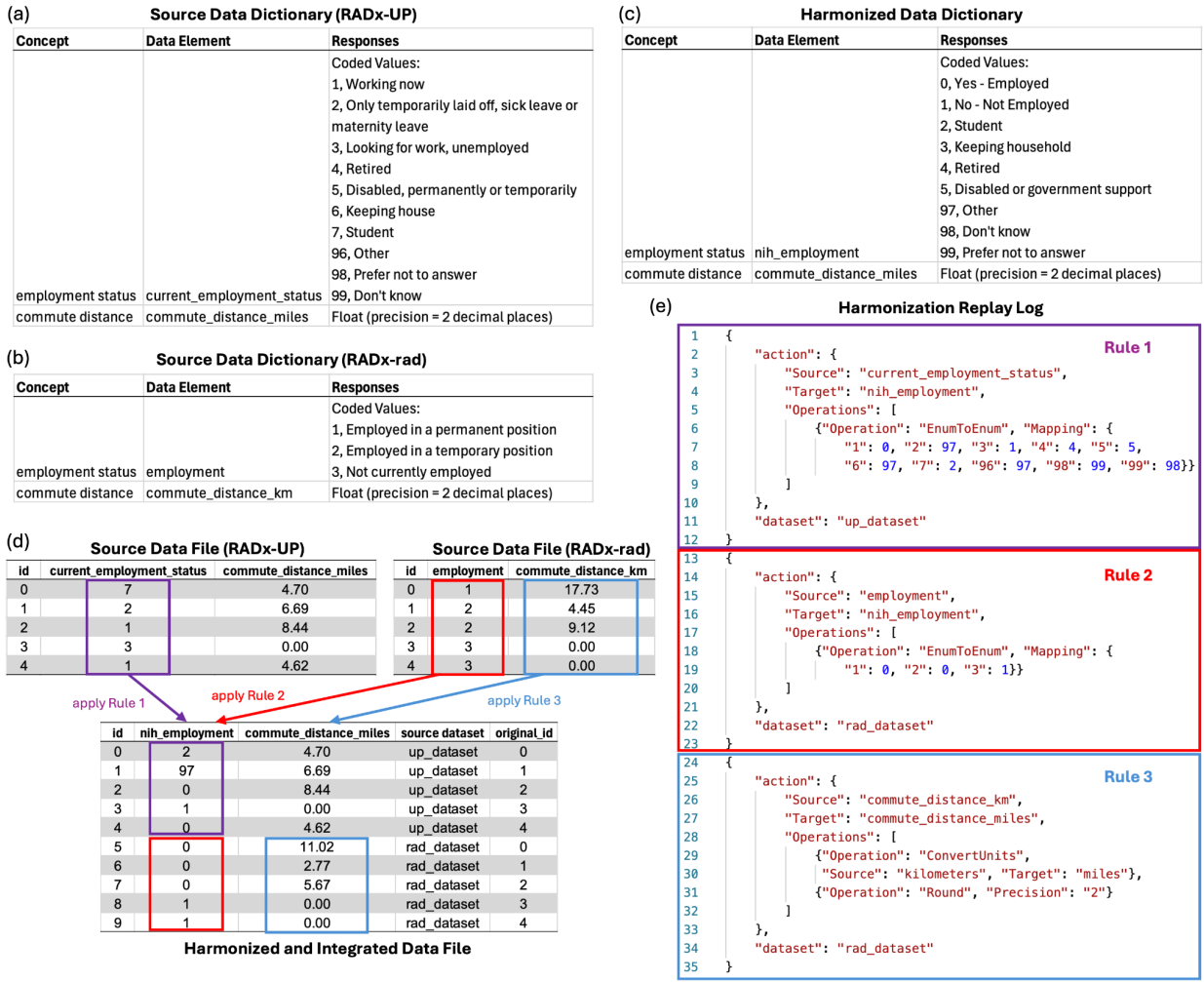

**Figure 6.** Application of our framework to synthetic data generated using data elements from RADx-UP and RADx-rad. (a) Data dictionary containing the "current_employment_status" data element from RADx-UP and the "commute_distance_miles" data element. (b) Data dictionary containing the "employment" data element from RADx-rad and the "commute_distance_km" data element. (c) Harmonized data dictionary containing target data elements "nih_employment" and "commute_distance_miles" used to harmonize the RADx-UP and RADx-rad datasets. (d) The harmonization framework ingests source data files corresponding to the data dictionaries in (a) and (b) and applies a set of user-defined harmonization rules to produce a harmonized and integrated dataset. The colors correspond to the subset of the data transformed by each harmonization rule, purple for Rule 1, red for Rule 2, and blue for Rule 3. (e) Replay log (in JSON format) for the harmonization process in (d). The log includes a serialization of the applied harmonization rule and names the dataset to which it was applied.

As part of data ingestion in the RADx Data Hub, data elements are harmonized to a set of CDEs marked by an

"nih_" prefix.[2] Data elements covering the "employment status" concept are harmonized to the "nih_employment" CDE. For our demonstration, we choose to standardize commute distances to be represented in units of miles. The target data dictionary for our harmonization process is shown in Figure 6c. The harmonization framework is used to implement mappings between the source data dictionaries in Figure 6a,b to the target data dictionary in Figure 6c in terms of the harmonization primitives listed in Table 1. Each mapping between data elements requires a separate harmonization rule. For the RADx-UP dataset, we require only a mapping from "current_employment_status" to "nih_employment" (Rule 1), as it already uses a data element for commute distance that is identical to the target. The RADx-rad dataset, however, requires harmonization across both of its data elements, from "employment" to "nih_employment" (Rule 2) and from "commute_distance_km" to "commute_distance_miles" (Rule 3). The implementation of each harmonization rule is shown in Figure 6e in its JSON serialized form. Rule 1 and Rule 2 are each implemented using only an *EnumToEnum* primitive that remaps coded variables, whereas Rule 3 requires a unit conversion from kilometers to miles and then rounding to match the desired output precision of two decimal places, using the *ConvertUnits* and *Round* primitives in sequence.

Figure 6d illustrates the application of each harmonization rule to the data files, with the color coding (purple for Rule 1, red for Rule 2, and blue for Rule 3) showing the blocks of data involved in each data transformation. The result is a harmonized and integrated dataset that complies with the target data dictionary in Figure 6c. Two auxiliary columns are introduced by the harmonization framework as part of this process. First, the *source dataset* column indicates the origin of the data. Second is the *original_id* column that records the index number in the original dataset for each entry. These record keeping features provide the transparency needed to trace the origin of transformed entries in the integrated and harmonized dataset, preventing future confusion following operations, such as sorting, that may shuffle the entries. Our framework supports scaling harmonization and integration protocols to many data files simultaneously, and the traceability of transformed data becomes increasingly important as the number and size of data files grow. The code that implements this demonstration is available on GitHub in the same repository as the framework software (see **Code Availability**).

**Discussion**
Our framework offers several notable advantages for data harmonization. It provides structure and machinery for implementing and executing harmonization protocols, which allows a data practitioner to rapidly prototype harmonization protocols and iterate on them as the need arises, such as when data collection templates change, when data standards evolve, or when new data types are introduced. The framework offers transparency by maintaining machine-readable harmonization rules and schema mappings to ensure that every data modification is explicitly logged and traceable. This capability affords clarity and consistency that may not be available in text-based, semi-automated, or manual protocols, allowing transformations to be reproduced identically across research groups or development environments. The common machinery of the framework also assists a user in scaling a harmonization protocol across many data elements and many data files as part of data integration work. Despite the advantages of leveraging our framework to perform harmonization, the formal approach introduces a learning curve, as writing ad hoc scripts for highly specific data transformations is much easier.

Several software tools for harmonization exist to support data harmonization across domains in biomedical informatics. Opal and Rmonize,[14] developed by Maelstrom Research, provide an integrated data management system to facilitate general-purpose harmonization protocols using the R programming language. However, both tools are coupled to comprehensive data management infrastructure and are primarily designed for data stewards and repository managers. HarmonizR is another R-based package tailored to omics data and focuses on handling batch effects and missing values, which ties its applicability closely to its domain.[15] DataHarmonizer is an interactive tool for curating tabular genomics data.[16] While it offers a user-friendly interface, it relies on a manual approach that may limit its ability to scale. Additionally, several works have proposed harmonization methodologies based on formal data element semantics[21–23] or on qualitative guidelines.[6,7] The formal approaches require significant expertise in formal data representations, whereas the qualitative approaches leave the implementation of harmonization machinery to the user.

Our harmonization framework is a lightweight general-purpose alternative to existing R-based harmonization software, which is especially important with Python having become the dominant programming language for data science applications. Our framework can accommodate data consumers, who are likely experts in their domain but not in data management, which is important when the goals of a data consumer diverge from those of a curator, as the semantically correct harmonization scheme may not be the most pragmatic alignment for a consumer of the data.

For example, a user may wish to reverse the mapping in Figure 5 to augment a dataset used for a secondary study on the correlations between age and some other variable to achieve higher statistical power. Such an application would require numerical age values rather than age ranges, but it is impossible to correctly map an age range to an integer age value. If the application tolerates large enough error bars, the user may decide that it suffices to transform a selected age range to any integer in the range or perhaps arbitrarily to the center value with some uniform jitter. Our framework supports such data transformations, only demanding that they be reproducible. This allows users to build pragmatic harmonization protocols, even when they are not semantically correct.

The distinguishing feature of our framework is its composition of primitive building blocks to implement harmonization rules. Although less formal compared to ontology-based methods, this approach is adaptable and extensible to modern machine learning workflows. Notably, large language models (LLMs) can be readily leveraged to generate harmonization rules using the available primitives as a foundation. We are investigating improvements in correctness and consistency through prompting structured by the set of primitive functions.

As part of our efforts to perform additional post hoc harmonization to enhance the interoperability and reusability of datasets currently hosted by the RADx Data Hub, we discovered a critical need for a general-purpose tool to facilitate data transformations and ensure their reproducibility, especially as a replacement for manual mappings and rigid transformation scripts that become obsolete as data standards evolve and research domains expand. As stewards of the RADx Data Hub, our applications of the harmonization framework focus on aligning ingested datasets against an established set of CDEs. However, we recognize that harmonization is a subjective process and that the RADx CDEs will not necessarily be the most suitable data representation for consumers of the data. Accordingly, we have made our framework accessible to any researchers, analysts, or data scientists who may require alternative data representations tailored to their research objectives. Our framework complements the Data Hub's integrated platform for data exploration and analysis by offering users the ability to harmonize and integrate RADx datasets as part of their workflows. Ultimately, the end goal of data harmonization is to make datasets useful to the consumer, and the availability of our framework offers that to all users of the RADx Data Hub.

**Conclusion**

We have presented a novel technique for data harmonization with an emphasis on the principled construction of harmonization rules and reproducibility with explicit record keeping. We have implemented a software framework as an open-source Python library and have provided a description of the models underlying our methodology and a proof-of-concept demonstration on synthetic data that align to data elements used by the RADx programs and Data Hub. Our software is already available for use, but there remain ongoing efforts to improve the accessibility of data harmonization to users of varying technical proficiency. Our framework currently supports only a library API, which requires users to write Python scripts or work in an interactive notebook. To eliminate this technical barrier, we are developing a graphical user interface that allows users to easily define harmonization transformations and execute them in a standalone application. Additionally, we are currently developing support for a more diverse set of harmonization procedures by expanding the available primitive functions to cover a larger range of data types. Notably, our proof-of-concept framework supports one-to-one data element mappings, but one-to-many, many-to-one, and many-to-many transformations will be necessary to expand the applicability of the framework.

Our framework introduces a new perspective on data harmonization that supports varied applications at multiple scales, from curating data for ingestion into permanent storage to rapid manipulation of data from multiple heterogeneous sources for data science applications. Although adoption of the framework introduces a learning curve, the high level of transparency, scalability, and reproducibility in harmonization protocols and harmonized datasets are valuable features for data curators and consumers alike.

**Acknowledgements**
This work was supported by the National Institutes of Health (NIH) under Other Transactions Authority award 1OT2DB000009-01.

**Code Availability**
The open-source library for the harmonization framework is available on GitHub under the BSD 2-Clause License at https://github.com/bmir-radx/harmonization-framework.